\documentclass[a4paper,11pt]{article}
\usepackage[scale={0.8,0.9},centering,includeheadfoot]{geometry}
\usepackage{amssymb,amsmath, graphicx,epsfig,amscd,array,color}
\usepackage{amsthm}
\usepackage[round]{natbib}

 \newcommand{\rINLA}{\texttt{r-INLA}\ }
 \def\mm#1{\ensuremath{\boldsymbol{#1}}} 

\usepackage{Sweave}
\begin{document}

\title{Fast approximate inference with INLA: the past, the present and the future  }
\author{Daniel Simpson, Finn Lindgren and H\aa{}vard Rue\\
    Department of Mathematical Sciences\\
    Norwegian University of Science and Technology\\
    N-7491 Trondheim, Norway} 
    \date{May 15, 2011}

\maketitle
\begin{abstract}
Latent Gaussian models are an extremely popular, flexible class of models. Bayesian inference for these models is, however, tricky and time consuming. Recently, Rue, Martino and Chopin introduced the Integrated Nested Laplace Approximation (INLA) method for deterministic fast approximate inference. In this paper, we outline the INLA approximation and its related R package. We will discuss the newer components of the r-INLA program as well as some possible extensions.
\end{abstract}

\section{Introduction}

As the statistical understanding of applied scientists increases and new techniques deliver larger, more complicated data sets, applied statisticians are faced with increasingly complex models.  Naturally, as the complexity of these models increase, it becomes harder and harder to perform inference. Appropriately, a great deal of effort has been expended on constructing numerical methods for performing approximate Bayesian inference. Undoubtably, the most popular family of approximate inference methods in Bayesian statistics is the class of Markov Chain Monte Carlo  (MCMC) methods.  These methods, which exploded into popularity in the mid 1980s and have remained at the forefront of Bayesian statistics ever since, with the basic framework being extended to cope with increasingly more complex problems.

The key advantage of MCMC methods is that, in their most vanilla incarnation, they are extremely simple to program.  This simplicity, together with their incredible flexibility, has lead to the proliferation of these methods.  Of course, there are problems: a single site auxiliary Gibbs sampler for spatial logistic regression is known to fail spectacularly.  This is just the tip of the iceberg---for even mildly complicated models, it can be extremely difficult to construct a MCMC scheme that converges in a reasonable amount of time.

For large models, and especially spatial models, fast convergence isn't enough.  Even if you could sample exactly from the posterior, sampling--based methods converge like $\mathcal{O}(N^{-1/2})$, where $N$ is the number of samples, which suggests that you need $10^{2p}$ samples to get an error of around $10^{-p}$.  Clearly, if computing a single sample is even reasonably expensive, this cost will be prohibitive.  In the best case, this means that MCMC schemes for large problems typically take hours or even days to deliver estimates that are only correct to three or four decimal places.  Clearly this is less than ideal!

The only way around this efficiency problem is to consider alternatives to sampling-based methods.  The first step in constructing an efficient approximate inference scheme is to greatly restrict the class of models that we will consider: it is na\"ive to expect that an efficient algorithm exists that will solve all of the problems that MCMC treats.  With this in mind, we restrict our attention to the class of \emph{latent Gaussian models}, which we define in three stages as 
\begin{align*}
y_i | \mathbf{x} &\sim \pi(y_i | x_i) \qquad &&\text{(Observation equation)} \\
\mathbf{x}|\boldsymbol{\theta} & \sim N(\boldsymbol{\mu}(\boldsymbol{\theta}),\mathbf{Q}(\boldsymbol{\theta})^{-1}) \qquad &&\text{(Latent Gaussian field)} \\
\boldsymbol{\theta} &\sim \pi(\boldsymbol{\theta}) \qquad &&\text{(Parameter model)},
\end{align*}
where $\mathbf{Q}(\boldsymbol{\theta})$ is the \emph{precision matrix} (that is, the inverse of the covariance matrix) of the Gaussian random vector $\mathbf{x}$.  In the interest of having a computable model, we will restrict $\mathbf{Q}$ to be either sparse or small enough that computing multiple factorisations is not an issue. These models cover a large chunk of classical statistical models, including dynamic linear models, stochastic volatility models, generalised linear (mixed) models, generalised additive (mixed) models, spline smoothing models, disease mapping, log-Gaussian Cox processes, model-based geostatistics, spatio-temporal models and survival analysis.

The Integrated Nested Laplace Approximation (INLA), builds upon the use of Laplace approximations, which were originally for approximating posterior distributions by \citet{art367}.  The first step in the INLA approximation is to perform a Laplace approximation to the joint posterior 
\begin{align}
\pi(\boldsymbol{\theta}|\mathbf{y}) &= \frac{\pi(\mm{\theta})\pi(\mm{x},\mm{\theta})\pi(\mm{y}|\mm{x})}{\pi(\mm{x}|\mm{\theta},\mm{y})} \notag \\
&\propto \frac{\pi(\mm{\theta})\pi(\mm{x},\mm{\theta})\pi(\mm{y}|\mm{x})}{\pi_G(\mm{x}|\mm{\theta},\mm{y})},  \label{laplace}
\end{align}
where $\pi_G(\mm{x}|\mm{\theta},\mm{y})$ is the Gaussian approximation to $\pi(\mm{x}|\mm{\theta},\mm{y})$ that matches the true distribution at the mode \citep{art451}.  The approximate posterior marginals for the non-Gaussian parameters can then be constructed through numerical integration as long as the dimension of $\mm{\theta}$ is not too large.  The posterior marginals for the latent field $\pi(x_i | \mm{y})$ are constructed by computing a Laplace approximation to $\pi(x_i | \mm{\theta},\mm{y})$ and then integrating out against the approximate joint posterior for $\mm{\theta}|\mm{y}$.  Full details of the approximation scheme can be found in \citet{art451}.

\section{The \rINLA program} \label{section2}
The INLA method was designed to be provide fast inference for a large class of practical Bayesian problems.  In order to fulfil this aim, the \rINLA package was created as an \texttt{R} interface to the INLA program, which is itself written in \texttt{C}.   The syntax for the \rINLA package is based on the inbuilt \texttt{glm} function in \texttt{R}, which highlights the effectiveness of the INLA method as a general solver for generalised linear (mixed) models.  The \rINLA package is available from \texttt{http://r-inla.org}.

They key to the computational efficiency of the \rINLA program is that it is based on \texttt{GMRFLib}, a \texttt{C} library written by H\aa{}vard Rue for performing efficient computations on Gaussian Markov random fields. As such, \rINLA is particularly effective when the latent Gaussian field has the Markov property.  This covers the case of spline smoothing (in any dimension), as well as conditional autoregressive models and some Mat\'{e}rn random fields \citep{Lindgren2011}. Such a latent field is specified through the \texttt{formula} mechanism in \texttt{R}.  

\begin{figure}[t]
  \centering

	\includegraphics[width=0.7\textwidth]{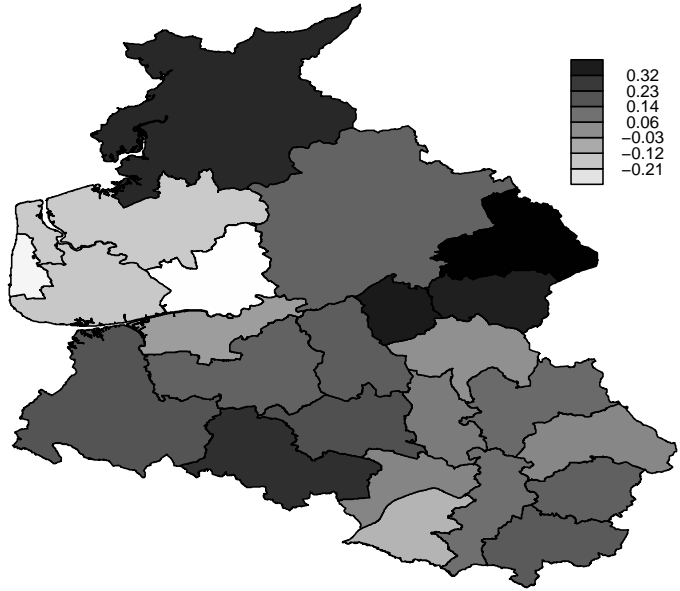}
\vspace{-2cm}
\caption{The posterior mean for the effect of district. \label{postmean}}
\end{figure}

To demonstrate the \rINLA package, let us consider some survival data for myeloid leukaemia cases in the north-west of England.  The model is a Cox proportional hazard model, where the hazard depends linearly on the age and sex of the patient, smoothly on the white blood count (\texttt{wbc}) and an econometric covariate (\texttt{tpi}).  Furthermore, it is assumed that there is a spatially correlated random effect, which takes into account which district the patient is in.  The following code performs full Bayesian inference on the appropriate generalised additive mixed model in around seven seconds.  The posterior mean spatial effect is shown in Figure \ref{postmean}
\begin{Schunk}
\begin{Sinput}
> data(Leuk)
> g = system.file("demodata/Leuk.graph", package = "INLA")
> formula = inla.surv(Leuk$time, Leuk$cens) ~ 1 + age + sex + f(inla.group(wbc), 
+     model = "rw1") + f(inla.group(tpi), model = "rw2") + f(district, 
+     model = "besag", graph.file = g)
> result = inla(formula, family = "coxph", data = Leuk)
\end{Sinput}
\end{Schunk}

\section{New features}
Since the original INLA paper, there have been a number of new developments.  In this section, we outline some of the most recent additions to the \rINLA package.

\paragraph{Manipulating the likelihood}

The original INLA method was limited to observation models where each observation depended on one element of the latent Gaussian field.  While this is commonly the case, this assumption is violated, for example, when the observed data consists of area averages of the latent field.  In this case, $$
y_i| \mm{x} \sim \pi\left(y_i \big| \sum_{j} a_{ij}x_j\right).
$$ We further assume that the dependence of the data on the latent field is ``local'' in the sense that most elements of the ``$\mm{A}$ matrix'' are zero.  With this assumption, everything stays Markovian and fast inference is still possible.  This is implemented in the \rINLA program by modifying the \texttt{control.compute} parameter in the \rINLA function call:
\begin{Schunk}
\begin{Sinput}
> res = inla(formula, family = "...", data = ..., control.compute = list(A = amat))
\end{Sinput}
\end{Schunk}

Beyond relaxing this restriction to the class of models considered by the \rINLA program, there are a number of other new methods for building new models.  The \texttt{f()} function, which \rINLA uses to specify random effects, has two new options: \texttt{replicate} and \texttt{copy}.  The first option can be used to simply deal with the case where the likelihood requires independent replicates of the model with the same hyperparameters.  The \texttt{copy}  option is useful in situations where the latent field uses the same random field multiple times, possibly with different scalings.  

Finally, \rINLA has been extended to include models where the data comes from different sources.  In this case, different subsets of the data will require different likelihood functions.  This can be programmed in \rINLA by re-writing the data as a matrix where the number of columns are equal to the number of likelihoods.  In the case where there are two likelihoods, each containing $n$ data points, this is achieved through the command 
\begin{Schunk}
\begin{Sinput}
> Y = matrix(NA, N, 2)
> Y[1:n, 1] = y[1:n]
> Y[1:n + n, 2] = y[(n + 1):(2 * n)]
\end{Sinput}
\end{Schunk}
The \rINLA command is then modified appropriately by setting \texttt{family = c("model1", "model2")}.

\paragraph{Survival models}
A class of models that were not considered in the original INLA paper were Bayesian survival models.  The trick is to see Bayesian survival models as just another set of Latent Gaussian models.  In some situations, this is straightforward, while at other times it requires data augmentation tricks, which are implemented in the \texttt{inla.surv()} function, demonstrated in Section \ref{section2}. These methods are outlined in \citep{tech90,tech91}, which also discuss ways to deal with different types of censoring. 

\paragraph{Stochastic partial differential equations}
A new method for constructing computationally efficient Gaussian random fields by taking advantage of the spatial Markov property was presented in a recent read paper by \citet{Lindgren2011}.   The idea is to use the fact that these fields can be represented as the solution to  stochastic partial differential equations (SPDEs) to construct computationally efficient approximations to them.  Beyond building computationally efficient approximations to standard spatial models, this method also allows for the construction of \emph{new} classes  of random fields with physically interpretable non-stationary.  These models have been implemented in \rINLA.  The following chunk of code fits a Bayesian spline through some noisy data points.  

It begins by constructing a mesh on the unit square with vertices at the observation locations (\texttt{points})
\begin{Schunk}
\begin{Sinput}
> bnd = inla.mesh.segment(matrix(c(0, 0, 1, 0, 1, 1, 0, 1), ncol = 2, 
+     byrow = TRUE))
> mesh = inla.mesh.create(points, boundary = bnd, refine = list(max.edge = 0.1))
\end{Sinput}
\end{Schunk}

The second step is to construct the SPDE model
\begin{Schunk}
\begin{Sinput}
> spde = inla.spde.create(mesh, model = "imatern")
\end{Sinput}
\end{Schunk}
where \texttt{imatern} is the intrinsic matern model with $\nu = 1$, i.e. the spline smoothing model. Finally the formula is constructed and the inference is performed in the standard way:
\begin{Schunk}
\begin{Sinput}
> formula = y ~ f(data_points, model = spde) - 1
> r = inla(formula, family = "gaussian", data = list(y = y, data_points = mesh$idx$loc))
\end{Sinput}
\end{Schunk}

\section{What the future holds}

There are an almost endless number of ways that the INLA method  \rINLA program can be extended.  In this section we describe some of the new features that we are currently working on.  

\paragraph{Fixing ``failures'':  global Gaussian approximations}

The Laplace approximation proceeds by fitting a Gaussian approximation around the mode of $\pi(\mm{x} | \mm{\theta},\mm{y})$, however there are situations in which this is not the most appropriate approximation.  For instance, if the true distribution is bimodal, a better `fit' would be obtained by constructing a Gaussian approximation that \emph{globally} approximates the distribution. 

\begin{figure}[t]
  \centering
	\includegraphics[width=0.6\textwidth]{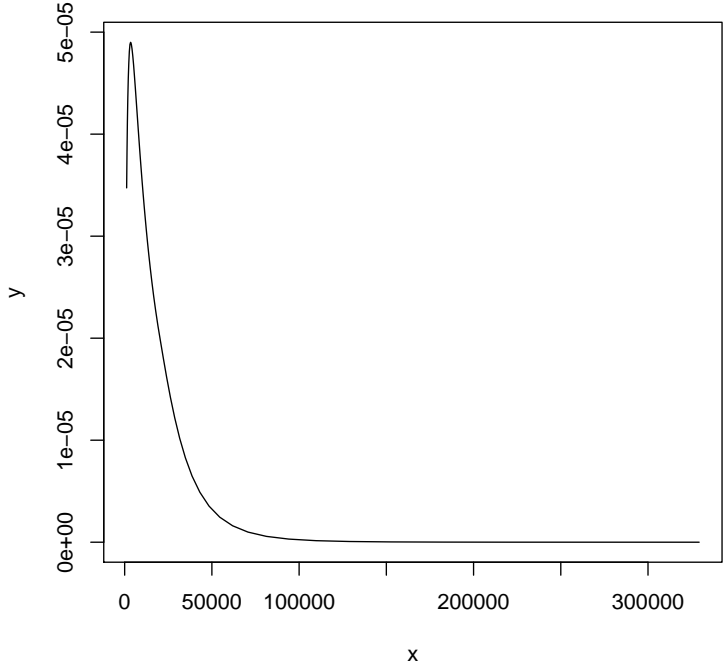}

\caption{The precision for the latent Gaussian field is badly overestimated---the true value is $\phi= 1$. \label{binomial_bad}}
\end{figure}

Another situation where these more global approximation would be of use is the following case of ``failure''.  Consider the problem of approximating the latent random field for the following logistic regression model.
\begin{Schunk}
\begin{Sinput}
> n = 100
> eta = 1 + rnorm(n)
> p = exp(eta)/(1 + exp(eta))
> y = rbinom(n, size = 1, prob = p)
> bad.result = inla(y ~ 1 + f(num, model = "iid"), family = "binomial", 
+     Ntrials = rep(1, n), data = list(y = y, num = c(1:100)))
\end{Sinput}
\end{Schunk}
Figure \ref{binomial_bad} shows the posterior for the precision of the random effect.  INLA has clearly missed the correct precision, which was $1$.  

So what went wrong?   Quite simply there is very little information in the data and hence the model is very prior sensitive.   This sensitivity, combined with the vague prior that \rINLA uses as a default produced the nonsense results in Figure \ref{binomial_bad}.

\paragraph{Kronecker product models}
In a number of applications, the precision matrix in the Gaussian random field can be written as a Kronecker product of two standard covariance matrices.  A simple example of this is the separable space-time model constructed by using spatially correlated innovations in an AR(1) model: $$
\mm{x}_{t+1} = \phi\mm{x}_t + \mm{\epsilon}_t,
$$ where $\phi$ is a scalar and $\mm{\epsilon} \sim N(0, {\mm{Q}_{\epsilon}}^{-1})$.   In this case, the precision matrix is $\mm{Q} = \mm{Q}_\text{AR(1)} \otimes \mm{Q}_{\epsilon}$, where $\otimes$ is the Kronecker product.

Due to the prevalence of Kronecker product models, it is desirable to add a Kronecker product mechanism to \rINLA.  The general Kronecker product mechanism is currently in progress, but a number of special cases are already available in the code through the undocumented \texttt{group} feature.  For example, a separable spatiotemporal SPDE model can be constructed using the command 
\begin{Schunk}
\begin{Sinput}
> frm = y ~ f(loc, model = spde, group = time, control.group = list(model = "ar1"))
\end{Sinput}
\end{Schunk}
in which every observation \texttt{y} is assigned a location \texttt{loc} and a time \texttt{time}.  At each time, the spatial points are linked by an SPDE model, while across the time periods, they evolve according to an AR(1) process. 

\paragraph{Extending the SPDE methodology}
The grouping mechanism described above can be used to produce separable space-time models, that is models in which the covariance function can be factored into a purely spatial and a purely temporal component. In some situations, this type of separability is an unrealistic assumption and a great deal of research has gone into constructing classes of non-separable spatiotemporal covariance functions.  An interesting property of SPDE models is that \emph{any} model built with a sensible space-time partial differential operator will lead to a non-separable model.  Furthermore, these models will inherit the good physical properties of the deterministic PDE models, such as causality and non-reversibility.  This guarantees that the non-separability is \emph{useful}, rather than simply present.

We are currently working to include the stochastic heat equation model $$
\frac{\partial }{\partial t} (\tau(s,t) x(s,t))  -\nabla\cdot \left(\mm{D}(s,t) \nabla(\tau(s,t)x(s,t))\right) + \nabla\cdot(\mm{b}(s,t)x(s,t)) + \kappa^2(s,t) x(s,t)= W(s,t),
$$ where the noise process  $W(s,t)$ is white in time, but correlated and Markovian in space.  The challenge here is not simply placing the model into the \rINLA framework. This model includes temporally varying anisotropy and temporally varying drift, and therefore, even parameterising this model is an open problem.  

\paragraph{Gamma frailty models: relaxing the Gaussian assumptions}

The assumption of Gaussian random effects is at the very heart of the INLA approximation.  However, there are a number of situations in which this is not a realistic assumption. An example of this comes when incorporating frailty into Cox proportional hazard models. In these models, the hazard function for individual $i$ is modelled as $$
h(t_i) = h_0 \nu_i \exp(\eta_i),
$$ where $\eta_i$ is a linear model containing covariates and $\nu_i$ is the frailty term, which models unobserved heterogeneity in the population.  Clearly, if we take $\nu_i$ to be log-normal, the resulting model fits firmly in the standard INLA framework.  Unfortunately, log-normal frailties are an uncommon model, typically the frailty term is taken to be gamma distributed.  The question is, therefore, can we incorporate gamma frailty models into the INLA framework.  

The solution to this problem comes in the guise of ``importance sampling''--type decomposition: 
$$
\text{Gamma} = \underbrace{\text{LogNormal}}_{\text{``Prior''}} \times \underbrace{\frac{\text{Gamma}}{\text{LogNormal}}}_\text{``Correction''}.
$$  With this type of formulation, it is possible to include gamma frailty models into the INLA framework.

This approach is not entirely satisfactory---although we can theoretically do this for any model suitably close to the log-normal (such as the log-t distribution), it is not particularly flexible.  The aim of this work is to incorporate ideas from Bayesian nonparametrics to construct a class of suitable non-Gaussian random effects models that can be incorporated into this framework.  This will massively increase the class of models for which INLA is available.

\section{Conclusion}
This article was finished on 15th May, 2011 and all of the information about INLA is correct at this time.  This statement is necessary---INLA is still a project in active development.  By the time you read this, some of the `present' features will have moved into the `past', and the `future' features will be edging ever closer to inclusion.    In fact, those who are interested can follow the progress of the INLA project at \texttt{http://inla.googlecode.com}, or by frequently updating the `testing' version of INLA using the command
\begin{verbatim}
> inla.update(testing=TRUE)
\end{verbatim}
This `testing' version of INLA updates frequently and includes experimental interfaces to the newest features.  This build also has the pleasant feature of matching with the documentation on \texttt{http://r-inla.org}!

The \texttt{r-INLA} project was created to provide an easy to use tool for performing Bayesian inference on latent Gaussian models.  As such, the set of problems that \texttt{r-INLA} can solve is limited to those that someone has wanted to solve.  There are any number of possible extensions not listed in the `future' section that we are not currently considering because no one has asked for them yet.  The lesson here is \emph{if you want \texttt{r-INLA} to have a particular feature, observation model or prior model, you need to ask us!} The development of the INLA project is driven entirely by the research interests of the development team and the requests that we receive from the user community.

\subsubsection*{Acknowledgements} 
The INLA team is H\aa{}vard Rue,  Sara Martino, Finn Lindgren, Daniel Simpson, Thiago Guerrera Martins and Rupali Akerkar.

\bibliographystyle{plainnat}
{\bibliography{mybib}}

\begin{thebibliography}{5}
\providecommand{\natexlab}[1]{#1}
\providecommand{\url}[1]{\texttt{#1}}
\expandafter\ifx\csname urlstyle\endcsname\relax
  \providecommand{\doi}[1]{doi: #1}\else
  \providecommand{\doi}{doi: \begingroup \urlstyle{rm}\Url}\fi

\bibitem[Akerkar et~al.(2010)Akerkar, Martino, and Rue]{tech90}
R.~Akerkar, S.~Martino, and H.~Rue.
\newblock Implementing approximate {B}ayesian inference for survival analysis
  using integrated nested {L}aplace approximations.
\newblock Technical report~1, Department of mathematical sciences, Norwegian
  University of Science and Technology, 2010.

\bibitem[Lindgren et~al.(2011)Lindgren, Lindstr{\"o}m, and Rue]{Lindgren2011}
F.~Lindgren, J.~Lindstr{\"o}m, and H.~Rue.
\newblock An explicit link between {G}aussian fields and {G}aussian {M}arkov
  random fields: {T}he {SPDE} approach.
\newblock \emph{Journal of the Royal Statistical Society. Series B. Statistical
  Methodology}, 2011.

\bibitem[Martino et~al.(2010)Martino, Akerkar, and Rue]{tech91}
S.~Martino, R.~Akerkar, and H.~Rue.
\newblock Approximate {B}ayesian inference for survival models.
\newblock Technical report~3, Department of mathematical sciences, Norwegian
  University of Science and Technology, 2010.

\bibitem[Rue et~al.(2009)Rue, Martino, and Chopin]{art451}
H.~Rue, S.~Martino, and N.~Chopin.
\newblock Approximate {B}ayesian inference for latent {G}aussian models using
  integrated nested {L}aplace approximations (with discussion).
\newblock \emph{Journal of the Royal Statistical Society, Series B},
  71\penalty0 (2):\penalty0 319--392, 2009.

\bibitem[Tierney and Kadane(1986)]{art367}
L.~Tierney and J.~B. Kadane.
\newblock Accurate approximations for posterior moments and marginal densities.
\newblock \emph{Journal of the American Statistical Association}, 81\penalty0
  (393):\penalty0 82--86, 1986.

\end{thebibliography}
\end{document}